\documentclass[conference]{IEEEtran}

\usepackage{graphicx}
\usepackage{mdframed}
\usepackage{xcolor}

\usepackage{url}
\usepackage{multirow}
\usepackage{subfigure}

\usepackage{tikz}
\usetikzlibrary{fit}
\usetikzlibrary{calc}
\usetikzlibrary{shapes}
\usetikzlibrary{positioning}
\usetikzlibrary{shadows}

\makeatletter
\pgfdeclareshape{document}{
\inheritsavedanchors[from=rectangle] 
\inheritanchorborder[from=rectangle]
\inheritanchor[from=rectangle]{center}
\inheritanchor[from=rectangle]{north}
\inheritanchor[from=rectangle]{south}
\inheritanchor[from=rectangle]{west}
\inheritanchor[from=rectangle]{east}
\backgroundpath{
\southwest \pgf@xa=\pgf@x \pgf@ya=\pgf@y
\northeast \pgf@xb=\pgf@x \pgf@yb=\pgf@y
\pgf@xc=\pgf@xb \advance\pgf@xc by-7.5pt 
\pgf@yc=\pgf@yb \advance\pgf@yc by-7.5pt
\pgfpathmoveto{\pgfpoint{\pgf@xa}{\pgf@ya}}
\pgfpathlineto{\pgfpoint{\pgf@xa}{\pgf@yb}}
\pgfpathlineto{\pgfpoint{\pgf@xc}{\pgf@yb}}
\pgfpathlineto{\pgfpoint{\pgf@xb}{\pgf@yc}}
\pgfpathlineto{\pgfpoint{\pgf@xb}{\pgf@ya}}
\pgfpathclose
\pgfpathmoveto{\pgfpoint{\pgf@xc}{\pgf@yb}}
\pgfpathlineto{\pgfpoint{\pgf@xc}{\pgf@yc}}
\pgfpathlineto{\pgfpoint{\pgf@xb}{\pgf@yc}}
\pgfpathlineto{\pgfpoint{\pgf@xc}{\pgf@yc}}
}
}
\makeatother

\newcommand{\am}{AMALTHEA}

\usepackage{caption}
\usepackage{subfigure}

\begin{document}

\title{A Workflow for Fast  Evaluation of Mapping Heuristics Targeting
  Cloud Infrastructures}

\author{
    \IEEEauthorblockN{Roman Ursu\IEEEauthorrefmark{1},
                      Khalid Latif\IEEEauthorrefmark{1},
                      David Novo\IEEEauthorrefmark{1},
                      Manuel Selva\IEEEauthorrefmark{1},
                      Abdoulaye Gamatie\IEEEauthorrefmark{1},
                      Gilles Sassatelli\IEEEauthorrefmark{1},
                    }
    \IEEEauthorblockN{Dmitry Khabi\IEEEauthorrefmark{2},
                      Alexey Cheptsov\IEEEauthorrefmark{2}
    }
    \IEEEauthorblockA{\IEEEauthorrefmark{1}LIRMM / CNRS - University of Montpellier, France
    \\\{firstname.lastname\}@lirmm.fr}
    \IEEEauthorblockA{\IEEEauthorrefmark{2}High-Performance Computing Center of Stuttgart (HLRS), Germany
    \\\{lastname\}@hlrs.de}
}
\maketitle

\begin{abstract}
  Resource   allocation   is  today   an   integral   part  of   cloud
  infrastructures management to  efficiently exploit resources.  Cloud
  infrastructures  centers generally  use custom  built heuristics  to
  define the resource allocations.  It is an immediate requirement for
  the management tools of these centers  to have a fast yet reasonably
  accurate simulation  and evaluation platform to  define the resource
  allocation for  cloud applications.  This work  proposes a framework
  allowing  users to  easily specify  mappings for  cloud applications
  described in the \am{} format used  in the context of the DreamCloud
  European project and to assess  the quality for these mappings.  The
  two quality metrics provided by the framework are execution time and
  energy consumption.
\end{abstract}

\IEEEpeerreviewmaketitle

\section{Introduction}

Usage  of cloud  infrastructures is  by no  means uncommon  in today's
scientific research.  A  typical cloud system is  a sophisticated tool
that relies  on the possibility  to combine computing,  networking and
storage  capabilities  of  multiple  compute resources  to  deliver  a
significant  performance advantage  over traditional  desktop systems.
Due   to  the   size,  cost   and  complexity   of  operation,   cloud
infrastructures  are  almost  never   used  exclusively  by  a  single
user. Instead, a  number of users try to compete  for access, and thus
mechanisms  for effective  resource sharing  among multiple  consumers
need to  be in place.   A process of allocating  resources efficiently
and scheduling computational tasks onto these allocated resources in a
proper  manner   is  based  on  resource   allocation  and  scheduling
priorities.   A typical  scheduler  targets  multiple goals  including
maximization of  throughput and  minimization of response  time (i.e.,
latency).   Other  optimization  criteria  such  as  CPU  fairness  or
energy-awareness  are becoming  more important  nowadays, too.   Since
some of  these goals  are conflicting  with each other,  a job  of the
scheduler  is to  figure out  a reasonable  allocation strategy.   The
approach of the DreamCloud project  is to enable a closed-loop control
mechanism that will allow for a dynamic resource allocation.
In order to assess the quality  of the resource allocation policy, the
use of dedicated fast simulators is of an essential importance.

{\bf Contributions of  this paper.}  To address  the above requirement
this  work  proposes a  configurable  framework  for fast  performance
evaluation of different resource allocation strategies targeting cloud
infrastructures.       The     implemented      framework     connects
\am{}~\cite{AMALTHEA} application models, used to describe application
as  a tasks  graph  in  the context  of  the  DreamCloud project  (see
Section~\ref{SEC:appmodel}  for  the  details  of the  model)  to  the
SimGrid~\cite{SimGrid}  simulation   platform.   It  aims  to   be  an
open-source and  flexible framework supporting  a variety of  state of
the art mapping algorithms.

{\bf    Outline.}    In    the    rest   of    the   paper,    Section
\ref{SEC:Related-Work}         discusses         related         work.
Section~\ref{SEC:appmodel} introduces  the \am{}  application modeling
language.  The  core engine  of our  simulation framework  allowing to
simulate  \am{}  applications  on  top  of  SimGrid  is  presented  in
Section~\ref{sec:simgrid2amalthea}.    The   proposed   framework   is
introduced  in Section~\ref{SEC:workflow}.   Section~\ref{SEC:results}
presents the  results of using  this flow  to evaluate the  quality of
different mappings for a scientific application case study executed on
top  of  a typical  cloud  infrastructure.   Finally, conclusions  and
perspectives are presented in Section \ref{SEC:Conclusions}.

\section{Related Work}
\label{SEC:Related-Work}

A  variety  of  cloud   application  simulation  frameworks  has  been
proposed.    CloudSim~\cite{cloudsim}   is   a   Java-based   software
framework, which supports modeling and simulation of large scale cloud
computing components and environments  including data centers, virtual
machines, service brokers, and  provisioning policies. The features of
CloudSim include the availability of  a visualization engine that aids
in the creation and management of multiple, independent, and co-hosted
virtualized services  on a  data center node.   Additionally, CloudSim
provides  flexibility   to  switch  between  space-   and  time-shared
allocation  of  processing  cores   to  virtualized  services.   These
compelling features of CloudSim would  speed up the development of new
application provisioning algorithms for  cloud computing.  However, it
is  not possible  to define  the  task level  allocation policies  for
CloudSim.

iCanCloud~\cite{iCanCloud}  is an  open  source  C++ based  simulation
platform which has been designed to model and simulate cloud computing
systems.  A  user can  customize the core  hypervisor class,  which in
turn is the  core of iCanCloud.  The Amazon model  has been integrated
with  the  simulator.   From   the  DreamCloud  perspective,  being  a
commercial tool, it does not  provide complete access to flexibly deal
with  application  requirements.   There   is  no  support  for  power
consumption  analysis.  Additionally,  the  focus of  iCanCloud is  to
define and integrate new brokering policies and minimize the user cost
for  public   cloud  infrastructure  based  on   classical  scheduling
heuristic. While the  main focus of the DreamCloud  project is dynamic
resource allocation.

GreenCloud~\cite{GreenCloud}  is  a sophisticated  packet-level  cloud
simulator with  focus on energy consumption  for cloud communications.
It  is  a   C++/OTcl  based  tool,  with   underlying  NS-2  platform.
GreenCloud provides  the fine-grained  modeling of  energy consumption
for  the IT  equipment in  a data  center, such  as computer  servers,
network  switches   and  communication  links.  From   the  DreamCloud
perspective, GreenCloud does not  provide the detailed timing analysis
which  is an  important  requirement for  evaluating dynamic  resource
allocations.  To  introduce the  detailed timing analysis,  a designer
needs  to define  timing models  in  NS2 environment,  which is  quite
outdated and very complex tool.

SimGrid~\cite{SimGrid} is  an open-source  and mature C-based  tool to
study the behavior of large-  scale distributed systems such as grids,
clouds, HPC  or P2P systems.  It  can be used to  evaluate heuristics,
prototype applications  or even assess legacy  MPI applications.  Task
level  scheduling  is supported  by  SimGrid.   The power  consumption
analysis  tool   has  been  integrated   to  the  latest   version  of
SimGrid. Additionally, the developers are very reactive.  Due to these
facts,  and making  the  basic  study, SimGrid  was  selected for  the
integration  to our  design  framework for  performance evaluation  of
applications executed on top of cloud infrastructures.

\section{Application Model}
\label{SEC:appmodel}
In  the   context  of  the  DreamCloud   project  addressing  resource
allocation issues  in future  multicore and distributed  systems. From
the applications  perspective, the project focuses  both on automotive
and on cloud computing domains.  The \am{}~\cite{AMALTHEA} application
model has been chosen to capture application specifications from these
two domains because it provides clean abstractions allowing to capture
both of  them.  \am{} is  based on  three main entities:  {\bf labels}
representing  data elements,  {\bf  runnables}  denoting the  smallest
units of code schedulable by an  OS, and {\bf tasks} defined as graphs
of runnables.

\begin{figure}[t]
  \centering
  \includegraphics[width=85mm]{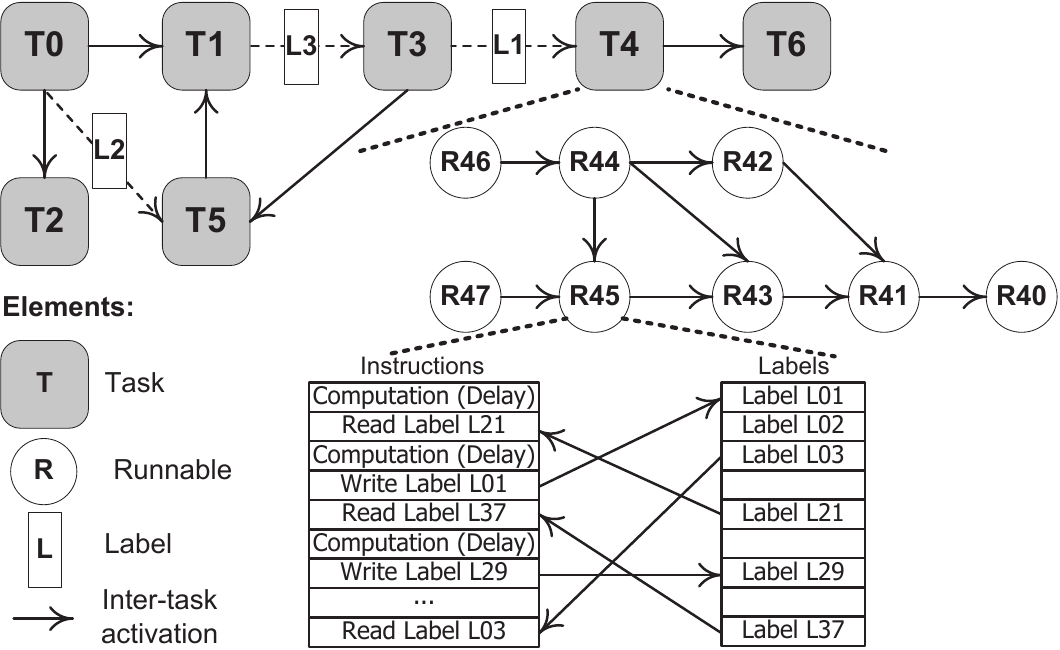}
  \caption{An illustrative representation of  an application as a task
    graph and task as a runnable graph, with inter-task activation and
    communication.}
  \label{FIG:AMALTHEA-SAMPLE}
\end{figure}

In  \am{}, an  application model  is  represented as  a directed  task
graph, where vertices represent  tasks, while directed edges represent
either    inter-task    activation     or    communication.     Figure
\ref{FIG:AMALTHEA-SAMPLE}  represents an  arbitrary \am{}  application
model with  seven tasks and  three labels.  To specify  the inter-task
communication, it can be observed  that task T1 communicates with task
T3 via label  L3. For this purpose, T1 will  perform a write operation
on L3 while T3 will perform the  read operation on L3.  The size of L3
represents the exchanged  data volume.  At lower  granularity level, a
task   is   composed   of   runnables   as   illustrated   in   Figure
\ref{FIG:AMALTHEA-SAMPLE}.  Task 'T4' is  composed of eight runnables.
The  runnable   graph  also  provides  information   about  precedence
relationship between runnables.

A runnable is a set of ``abstract'' instructions. Such instructions in
\am{}   application  model   are   categorized   as  computation   and
communication   instructions.     Computation   instructions   feature
operations  such  as  arithmetic  or logical  operations.   For  quick
simulations, the actual computational instructions are abstracted away
by representing them with associated platform-dependent execution time
in  terms  of  clock   cycles.   Communication  instructions  comprise
read/write accesses  to labels, which  can be further  classified into
two categories: local and remote  accesses. Local accesses take place,
when the  target label is  mapped at the  same node as  the requesting
runnable. Remote accesses take place,  when the target label is mapped
on a different  node, compared to that of the  requesting runnable. In
this  case, the  runnable  goes to  the  wait state  to  wait for  the
read/write response and  makes the core available  for other runnables
to  execute  their   instructions.   Figure  \ref{FIG:AMALTHEA-SAMPLE}
details the  instructions of runnable  R45 from the runnable  graph of
task  T4.  The  computation  instructions are  represented with  their
execution   delay   values.    The  communication   instructions   are
represented   as   read/write    accesses   to   their   corresponding
labels. Their simulation consists of  the actual transfer of data size
corresponding that of accessed labels in read/write requests.

\section{From SimGrid to \am{}}
\label{sec:simgrid2amalthea}

We now  review how we  implemented the simulation  of \am{} on  top of
SimGrid.   Before digging  into  the details,  we  describe the  basic
architecture of SimGrid.

\subsection{SimGrid Architecture}
\label{sec:simgrid-architecture}

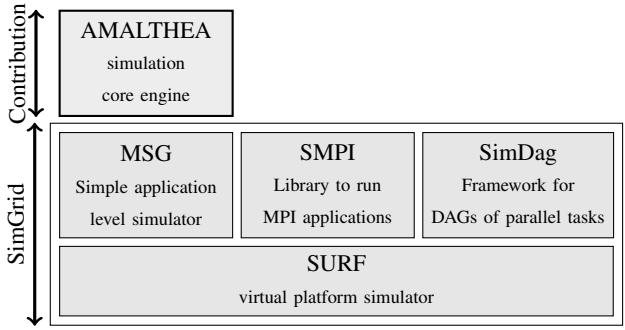
\begin{figure}[t]
  \centering
  \begin{tikzpicture}
        \tikzstyle{block}=[align=center, fill=black!10, draw=black, minimum width=23mm]
        \node[block, minimum height=14mm] (MSG) {\small MSG\\ \scriptsize Simple application\\ \scriptsize level simulator};
        \node[block, minimum height=14mm, anchor=west, xshift=1mm] (SMPI) at (MSG.east) {\small SMPI\\ \scriptsize Library to run \\ \scriptsize MPI applications};
        \node[block, minimum height=14mm, anchor=west, xshift=1mm] (SIMDAG) at (SMPI.east) {\small SimDag\\ \scriptsize Framework for \\ \scriptsize DAGs of parallel tasks};
        \node[block, minimum width=73.6mm, anchor=north west, yshift=-1mm] (surf) at (MSG.south west) {\small SURF\\ \scriptsize virtual platform simulator};
        \node[fit=(MSG)(surf), draw=black] (fit) {};
        \node[block, fill=black!7, thick, minimum height=14mm, anchor=south, yshift=2mm] at (MSG.north) (contrib) {\small \am{} \\ \scriptsize simulation\\ \scriptsize core engine};
        \draw[<->, very thick] ($(fit.north west) - (2mm,0)$) -- node[rotate=90, above]{\small SimGrid} ($(fit.south west) - (2mm,0)$);
        \draw[<->, very thick] ($(contrib.north west) - (3mm,0)$) -- node[rotate=90, above]{\small Contribution} ($(contrib.south west) - (3mm,0)$);

  \end{tikzpicture}
  \caption{SimGrid   Architecture  and   the   contribution  of   this
    work. SimGrid provides three different  APIs allowing user code to
    write custom simulations.  This work relies on the MSG API.}
  \label{simgrid-fig}
\end{figure}

SimGrid has been an active project for more than 10 years. Since then,
it has  been evolving  and taking new  systems into  consideration. At
start, it was dedicated to grid  simulation and its focus was to study
centralized  scheduling simulations  (e.g., off-line  scheduling of  a
Directed Acyclic  Graph (DAG)  on a  heterogeneous set  of distributed
computing  nodes).   Nowadays,  SimGrid   deals  with  a   variety  of
distributed systems  ranging from  grids to peer-to-peer  systems, and
arbitrarily  defined  hybrid  systems  combining  different  computing
systems. The latest, stable and  publicly available version of SimGrid
(3.2) implements the architecture shown in Figure~\ref{simgrid-fig}.

SimGrid provides  three APIs, all  based on a common  kernel, allowing
users to write  custom simulations. SURF is this  internal kernel.  It
provides the core functionalities to  simulate a virtual platform.  It
has been implemented at very low level  and is not intended to be used
by the  end users.  On  top of the SURF  layer, SimDag, MSG,  and SMPI
APIs are implemented.  MSG is  a simple programming environment, which
was  the first  distributed  programming  environment provided  within
SimGrid.  It is  still the most commonly  used programming environment
and can be used to build rather realistic simulations.  The SimDag API
provides specific  programming environment for DAG  applications while
the SMPI  one provides  support for  simulating MPI  applications.  To
simulate  the rich  set of  abstractions provided  by the  \am{} model
described in the previous section, we rely on the MSG API.

\subsection{\am{} Execution on SimGrid}
\label{sec:am2simgrid-exec}

To execute \am{}  applications on top of SimGrid,  we implemented a
simulation core engine that  reads the in-memory object representation
of an \am{} model and converts  it to SimGrid function calls.  Each
runnable  of  the initial  application  is  converted into  a  SimGrid
process.  Runnables activation dependencies and communications through
labels are then implemented using SimGrid messages.

In order to fit \am{}  applications into SimGrid, we simplified the
\am{}  semantic  by  gathering  for  each  runnable  all  the  read
instructions together, all the computing instructions together and all
the  write instructions  together.  From  that, each  runnable process
first waits for  receiving activation messages from  all the runnables
it   depends  on.    Then,   the  runnable   process  sends   messages
corresponding  to  the  read  label  instructions  it  contains.   The
computational load of the process required by SimGrid is then computed
by adding  the execution time of  all the compute instructions  of the
runnable.   Finally,   the  runnable  process  performs   write  label
operations and activates all subsequent runnables.

\section{Workflow}
\label{SEC:workflow}

In order to be able to  simulate \am{} applications on top of cloud
infrastructure    we     propose    the    workflow     depicted    on
Figure~\ref{workflow-fig}.  This workflow  comprises  the 4  following
steps: \vspace{0.2cm}

\begin{figure}[t]
  \centering
  \begin{tikzpicture}
    \tikzstyle{io}=[rounded corners=0.2cm, align=center, fill=black!30, draw=black]
    \tikzstyle{block}=[align=center, fill=black!10, draw=black, minimum width=45mm]
    \node[io] (in) {\small Application\\ \small specifications};
    \node[block, anchor=north, yshift=-3mm] (amalthea) at (in.south) {\small \am{} environment};
    \node[anchor=west] at (amalthea.west) {\small \textbf{1}};
    \node[block, anchor=north, yshift=-3mm] (parser) at (amalthea.south) {\small Parser};
    \node[anchor=west] at (parser.west) {\small \textbf{2}};
    \node[block, anchor=north, yshift=-3mm] (mapper) at (parser.south) {\small Mapper};
    \node[anchor=west] at (mapper.west) {\small \textbf{3}};
    \node[block, anchor=north, yshift=-3mm] (sim) at (mapper.south) {\small \am{} simulator \\ \scriptsize Section~\ref{sec:am2simgrid-exec}};
    \node[io, anchor=west, inner sep=2pt, xshift=4mm] (archi) at (sim.north east) {\small Archi.\\\small description\\ \small (xml)};
    \node[anchor=west] at (sim.west) {\small \textbf{4}};
    \node[
    shape=document,
    double copy shadow={
      shadow xshift=-0.5ex,
      shadow yshift=-0.5ex
    },
    draw,
    fill=white,
    line width=1pt,
    text width=0.4cm,
    minimum height=0.9cm, anchor=north, xshift=-8mm, yshift=-3.5mm] (out) at (sim.south) {};
    \node[align=center, font=\tiny] at (out) {Enery\\Time};

    \draw[->, thick] ($(sim.south) + (5mm,-13mm)$) -- ($(sim.south) + (5mm,-5mm)$);
    \draw[->, thick] ($(sim.south) + (5mm,-13mm)$) -- ($(sim.south) + (15mm,-13mm)$);
    \node[anchor=north west, xshift=4mm] at ($(sim.south) + (5mm,-13mm)$) {\scriptsize time};
    \node[anchor=south, xshift=0mm] at ($(sim.south) + (5mm,-6mm)$) {\scriptsize hosts};
    \node[draw=black, fill=black!5, minimum width=4mm, , inner sep=2pt] at ($(sim.south) + (8mm,-7.5mm)$) {};
    \node[draw=black, fill=black!5, minimum width=5.5mm, inner sep=2pt] at ($(sim.south) + (10mm,-9.5mm)$) {};
    \node[draw=black, fill=black!5, minimum width=2.5mm, inner sep=2pt] at ($(sim.south) + (7mm,-11.5mm)$) {};

    \draw[->, very thick] (in) -- (amalthea);
    \draw[->, very thick] (amalthea) -- (parser);
    \draw[->, very thick] (parser) -- (mapper);
    \draw[->, very thick] (mapper) -- (sim);
    \draw[->, very thick] (sim.south) -- ++ (0,-3mm);
    \draw[-, very thick] (archi.west) -- ++(-2mm,0);
    \draw[->, very thick] ($(archi.west) +(-2mm,0)$) |- (mapper.east);
    \draw[->, very thick] ($(archi.west) +(-2mm,0)$) |- (sim.east);

  \end{tikzpicture}
  \caption{Simulation  workflow  from applications  specifications  to
    simulation results}
  \label{workflow-fig}
\end{figure}
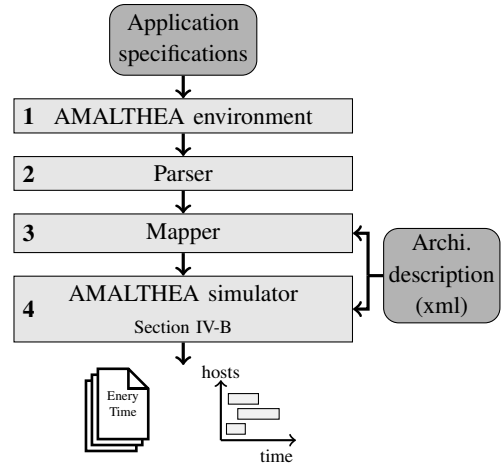

\begin{itemize}
\item {\bf  Step 1.}   The application  specifications are  modeled in
  \am{} as discussed in the previous section.  \vspace{0.2cm}

\item {\bf Step 2.}  The \am{} model is fed to a parser. The job of
  the parser is to translate the \am{} model into a format that can
  then be easily simulated using SimGrid. This parser has been written
  in C++ to be easily integrated with SimGrid.
\vspace{0.2cm}

\item {\bf Step  3.}  The mapper then executes  the mapping algorithms
  by allocating  tasks and  labels onto  the simulation  platform. The
  description of the platform is provided to the mapper.  Because this
  work focuses  on providing  a modular  workflow to  evaluate mapping
  strategies,  the  proposed  workflow  does  not  provide  elaborated
  mapping strategies  but rather a  clean and simple interface  to let
  users implement any mapping  algorithm.  Nevertheless, to illustrate
  the  complete workflow,  basic  mapping strategies  are proposed  by
  default.
\vspace{0.2cm}

\item  {\bf  Step 4.}   At  this  stage,  the \am{}  runnables  are
  executed   on    the   simulation    platform   as    described   in
  Section~\ref{sec:am2simgrid-exec}.  Ensuring  the \am{} execution
  semantic  on   top  of   the  SimGrid  MSG   API  was   the  biggest
  implementation challenge of the workflow.
\end{itemize}

As  results,  the  simulation   platform  provides  information  about
execution time, execution steps and  the energy consumption. All these
results  are  provided  in  the   form  of  text  files.   The  energy
consumption  results   are  provided  both  globally   and  per  host.
Moreover, the workflow provides a  timeline view of runnable processes
execution on the different hosts of the platform.  This timeline view,
provided as  a standard timeline  trace file, also  includes runnables
dependencies and labels reading and writing information. This timeline
view is  provided as a trace  file that could be  visualized using the
Vite\footnote{http://vite.gforge.inria.fr} open source tool.

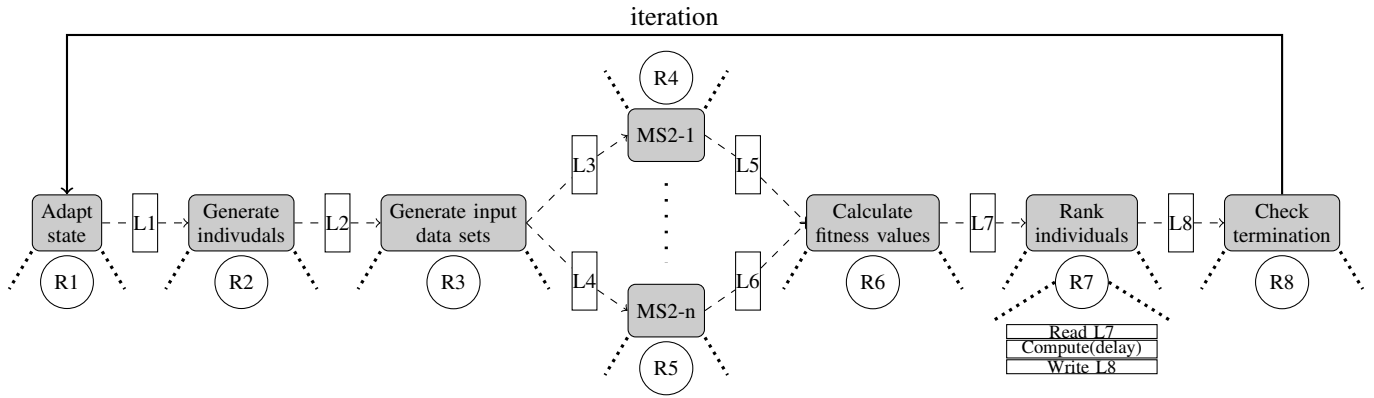
\begin{figure*}[t]
  \centering
  \begin{tikzpicture}[node distance=0mm and 4mm]
    \tikzstyle{task}=[font=\footnotesize,align=center, fill=black!20, draw=black, rounded corners=0.1cm]
    \tikzstyle{run}=[font=\footnotesize,align=center, circle, draw=black]
    \tikzstyle{label}=[font=\footnotesize,align=center, draw=black, minimum height=22, inner sep=0.25pt]

    \node[task] (adapt) {Adapt\\state};
    \draw[dotted, very thick] (adapt.south west) -- ++ (-3mm,-5mm);
    \draw[dotted, very thick] (adapt.south east) -- ++ (+3mm,-5mm);
    \node[run, below=of adapt, yshift=-0.5mm] (r1) {R1};

    \node[label, right=of adapt] (l1) {L1};

    \node[task, right=of l1] (genindiv) {Generate\\indivudals};
    \draw[dotted, very thick] (genindiv.south west) -- ++ (-3mm,-5mm);
    \draw[dotted, very thick] (genindiv.south east) -- ++ (+3mm,-5mm);
    \node[run, below=of genindiv, yshift=-0.5mm] (r2) {R2};

    \node[label, right=of genindiv] (l2) {L2};

    \node[task, right=of l2] (geninp) {Generate input\\data sets};
    \draw[dotted, very thick] (geninp.south west) -- ++ (-3mm,-5mm);
    \draw[dotted, very thick] (geninp.south east) -- ++ (+3mm,-5mm);
    \node[run, below=of geninp, yshift=-0.5mm] (r3) {R3};

    \node[label, above right=of geninp, xshift=2mm] (l3) {L3};
    \node[label, below right=of geninp, xshift=2mm] (l4) {L4};

    \node[task, above right=of l3, anchor=west, minimum height=7mm] (ms21) {MS2-1};
    \draw[dotted, very thick] (ms21.north west) -- ++ (-3mm,+5mm);
    \draw[dotted, very thick] (ms21.north east) -- ++ (+3mm,+5mm);
    \node[run, above=of ms21, yshift=+0.5mm] (r4) {R4};

    \node[task, below right=of l4, anchor=west, minimum height=7mm] (ms2n) {MS2-n};
    \draw[dotted, very thick] (ms2n.south west) -- ++ (-3mm,-5mm);
    \draw[dotted, very thick] (ms2n.south east) -- ++ (+3mm,-5mm);
    \node[run, below=of ms2n, yshift=-0.5mm] (r5) {R5};

    \node[label, below right=of ms21.east] (l5) {L5};
    \node[label, above right=of ms2n.east] (l6) {L6};

    \node[task, above right=of l6, xshift=2mm] (fitness) {Calculate\\fitness values};
    \draw[dotted, very thick] (fitness.south west) -- ++ (-3mm,-5mm);
    \draw[dotted, very thick] (fitness.south east) -- ++ (+3mm,-5mm);
    \node[run, below=of fitness, yshift=-0.5mm] (r6) {R6};

    \node[label, right=of fitness] (l7) {L7};

    \node[task, right=of l7] (rank) {Rank\\individuals};
    \draw[dotted, very thick] (rank.south west) -- ++ (-3mm,-5mm);
    \draw[dotted, very thick] (rank.south east) -- ++ (+3mm,-5mm);
    \node[run, below=of rank, yshift=-0.5mm] (r7) {R7};
    \draw[dotted, very thick] (r7.west) -- ++ (-8mm,-5mm);
    \draw[dotted, very thick] (r7.east) -- ++ (+8mm,-5mm);
    \node[draw=black, anchor=north, yshift=-2.00mm, inner sep=0.2pt, minimum width=20mm] (rd)  at (r7.south) {\scriptsize Read L7};
    \node[draw=black, anchor=north, yshift=-0.1mm, inner sep=0.2pt, minimum width=20mm] (cpt) at (rd.south) {\scriptsize Compute(delay)};
    \node[draw=black, anchor=north, yshift=-0.1mm, inner sep=0.2pt, minimum width=20mm] (cpt) at (cpt.south) {\scriptsize Write L8};

    \node[label, right=of rank] (l8) {L8};

    \node[task, right=of l8] (end) {Check\\termination};
    \draw[dotted, very thick] (end.south west) -- ++ (-3mm,-5mm);
    \draw[dotted, very thick] (end.south east) -- ++ (+3mm,-5mm);
    \node[run, below=of end, yshift=-0.5mm] (r8) {R8};

    \draw[dashed] (adapt) -- (l1);
    \draw[dashed, ->] (l1) -- (genindiv);
    \draw[dashed] (genindiv) -- (l2);
    \draw[dashed, ->] (l2) -- (geninp);
    \draw[dashed] (geninp.east) -- (l3);
    \draw[dashed, ->] (l3) -- (ms21.west);
    \draw[dashed] (geninp.east) -- (l4);
    \draw[dashed, ->] (l4) -- (ms2n.west);
    \draw[dash pattern=on 1pt off 5pt, very thick, shorten <=8pt, shorten >=8pt] (ms21) -- (ms2n);
    \draw[dashed] (ms21.east) -- (l5);
    \draw[dashed, ->] (l5) -- (fitness.west);
    \draw[dashed] (ms2n.east) -- (l6);
    \draw[dashed, ->] (l6) -- (fitness.west);
    \draw[dashed] (fitness) -- (l7);
    \draw[dashed, ->] (l7) -- (rank);
    \draw[dashed] (rank) -- (l8);
    \draw[dashed, ->] (l8) -- (end);

    \coordinate (abv) at (0,25mm);
    \draw[thick, ->] (end.north) -- (end.north |- abv) -- node[above] {iteration} (adapt.north |- abv) -- (adapt.north);

  \end{tikzpicture}
  \caption{The eScience case study application}
  \label{app-fig}
\end{figure*}

\section{Experimental Results}
\label{SEC:results}

We  now show  how the  proposed workflow  can be  used to  evaluate an
application  model and  a cloud  infrastructure provided  by the  High
Performance  Computing   Center  Stuttgart  (HLRS)  involved   in  the
DreamCloud project.   In all the following  experiments, to illustrate
the workflow,  we use a  mapping algorithm that randomly  allocate the
runnables and the labels on the underlying hardware infrastructure.

\subsection{eScience Application}\label{subsection:EScience Application}

The  application use  case is  based  on an  eScience application  MS2
genetic algorithm~\cite{Glass20143302} that performs molecular dynamic
simulation.  The goal  of the simulation is  to predict thermophysical
properties of  condensed fluids which  is required for the  design and
optimization of chemical engineering processes.  The eScience workflow
as an \am{} model is  depicted in Figure~\ref{app-fig}. MS2 is used
to  optimize the  algorithm that  tries to  fit the  parameters of  an
existing model  to data collected through  experiments.  The algorithm
is executed until a good enough fit is found, a time limit is hit or a
fixed number of iterations have been executed.  This in handled by the
task called \emph{Check termination}.  In this application model, each
\am{}   task   contains   a   single  runnable   as   depicted   on
Figure~\ref{app-fig}. As  shown for  the runnable  $R7$, each  of this
runnable first  reads input data from  an input label, then  perform a
specific number of computations  instructions and finally write output
data  to  an output  label.   In  all  the following  experiments,  32
\emph{MS2} tasks are considered.

\subsection{Cloud Infrastructure}

\begin{figure}[t]
  \centering
  \includegraphics[width=87mm]{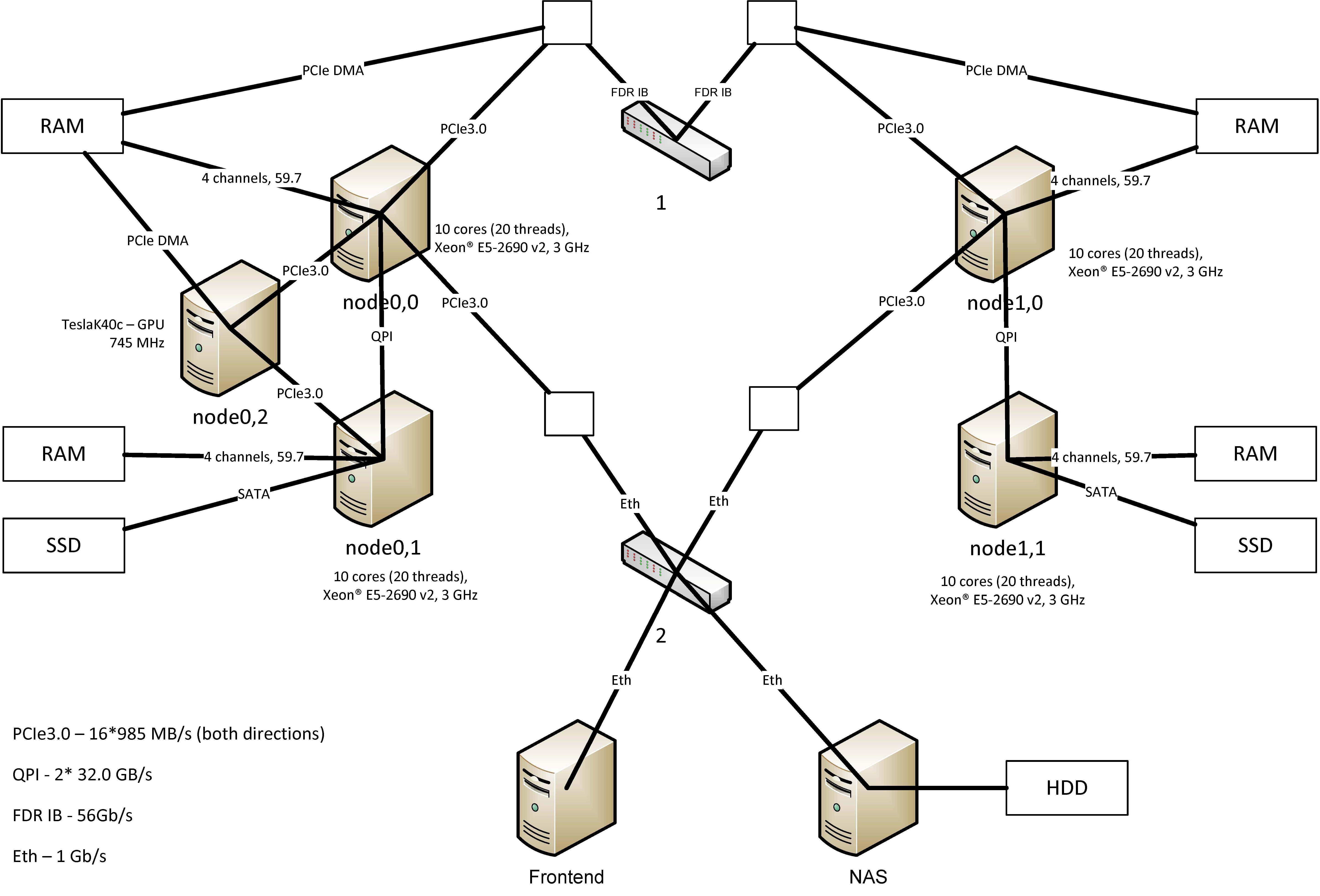}
  \caption{The HLRS cloud platform}
  \label{platform-fig}
\end{figure}

The cloud  infrastructure model we  use in the  following experiments,
also  provided by  HLRS,  is shown  in Figure~\ref{platform-fig}.   It
contains two  nodes, named node 0  and node 1.  Node  0 contains three
hosts (node0,0  node0,1 and node0,2)  while node 1 contains  two hosts
(node1,0 and node1,1). The configuration  details for each host can be
observed.   The  Frontend  host  acts  as  the  master,  and  resource
allocations are  performed by  this host.  The  bandwidth requirements
for each interconnection link have also been provided.

\subsection{Results}

\subsubsection{Preliminary mapping evaluation}

We simulate  the execution  of the eScience  application on  the cloud
infrastructure described previously.  During each simulation instance,
we trigger only one run of  the application, i.e. the execution of the
application  completes   as  soon   as  the  task   named  \emph{Check
  termination}  finishes. In  a  more general  execution setting,  the
eScience application  could possibly iterate several  times until data
fit the model (see Section \ref{subsection:EScience Application}).

On  the   other  hand,   we  consider  two   variants  of   the  cloud
infrastructure: heterogeneous  and homogeneous platform  versions. The
former is  depicted in Figure~\ref{platform-fig}, while  the latter is
obtained  by  removing  from   the  Heterogeneous  platform  the  host
HOST\_0\_2, which is a GPU.

\begin{figure}[t]
  \begin{center}
    \subfigure[Execution time]{\label{FIG:wgpu1}\resizebox{80mm}{!}{\includegraphics{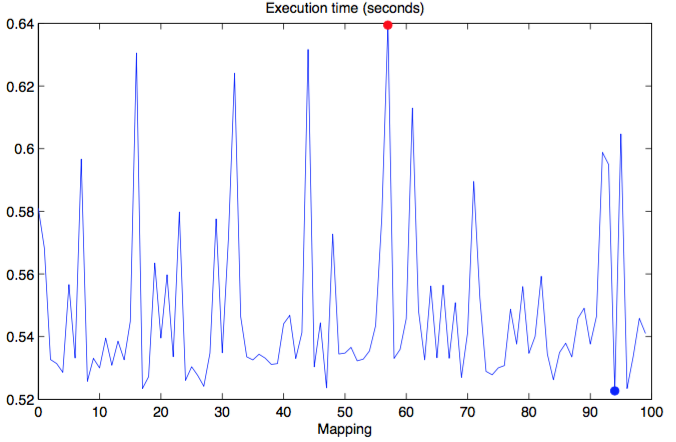}}} \qquad\qquad
    \subfigure[Energy consumption]{\label{FIG:wgpu2}\resizebox{80mm}{!}{\includegraphics{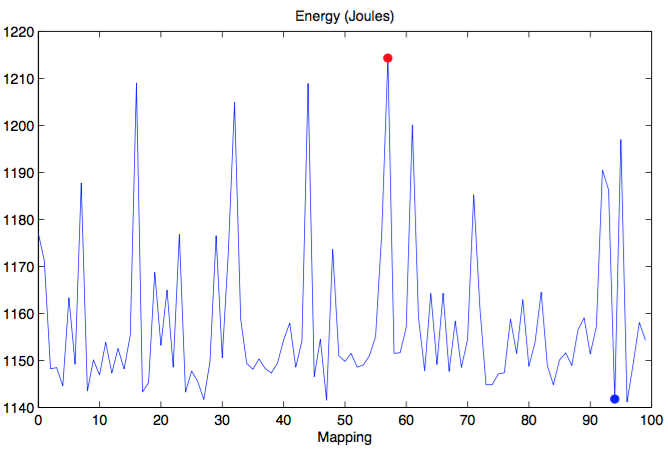}}}  \qquad\qquad
    \subfigure[Simulation time]{\label{FIG:wgpu3}\resizebox{80mm}{!}{\includegraphics{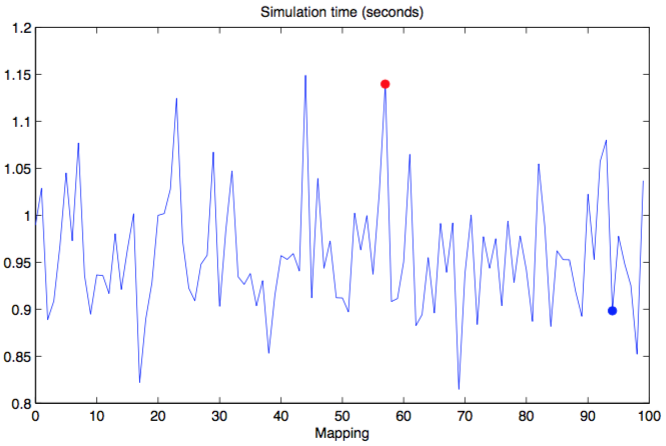}}}
  \end{center}
  \caption{Simulation on homogeneous platform}
  \label{FIG:homogeneous}
\end{figure}

We  conducted two  sets of  experiments. In  the first  experiment, we
simulated the execution of 100  random static mappings of the eScience
application on the homogeneous  platform. The characterization of each
mapping in terms of execution  time, energy dissipation and simulation
time is represented in  Figure~\ref{FIG:homogeneous}. The red and blue
dots in the Figure,  point to the worst and best  mappings in terms of
the execution time.

\begin{figure}[!t]
  \begin{center}
    \subfigure[Execution time]{\label{FIG:gpu1}\resizebox{80mm}{!}{\includegraphics{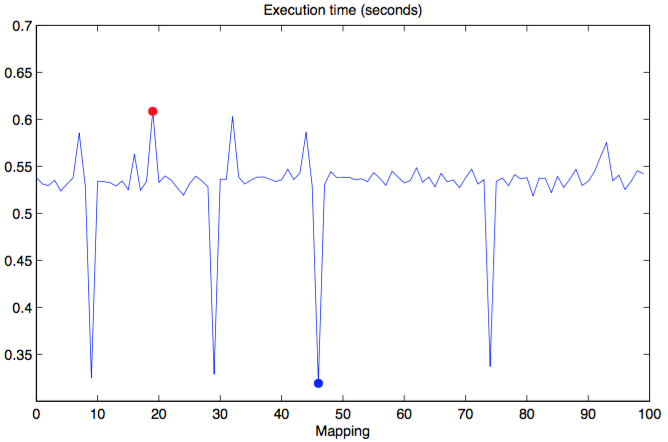}}} \qquad\qquad
    \subfigure[Energy consumption]{\label{FIG:gpu2}\resizebox{80mm}{!}{\includegraphics{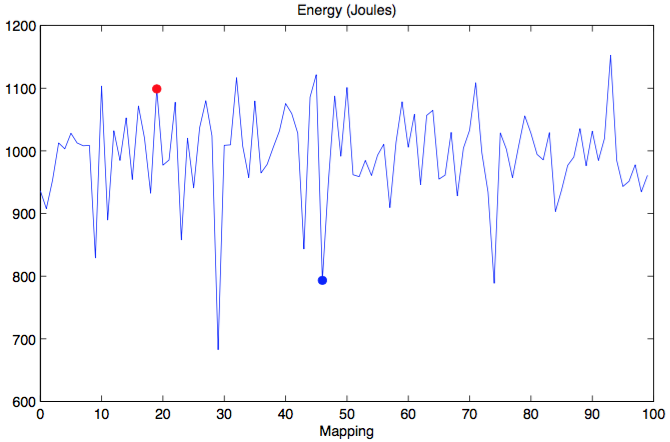}}}  \qquad\qquad
    \subfigure[Simulation time]{\label{FIG:gpu3}\resizebox{80mm}{!}{\includegraphics{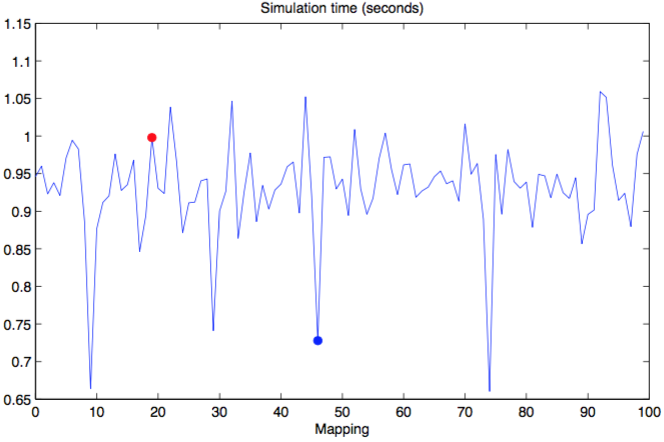}}}
  \end{center}
  \caption{Simulation on heterogeneous platform}
  \label{FIG:heterogeneous}
\end{figure}

\begin{figure*}[htbp]
  \begin{center}
    \subfigure[Best mapping scenario]{\label{FIG:best}\resizebox{150mm}{!}{\includegraphics{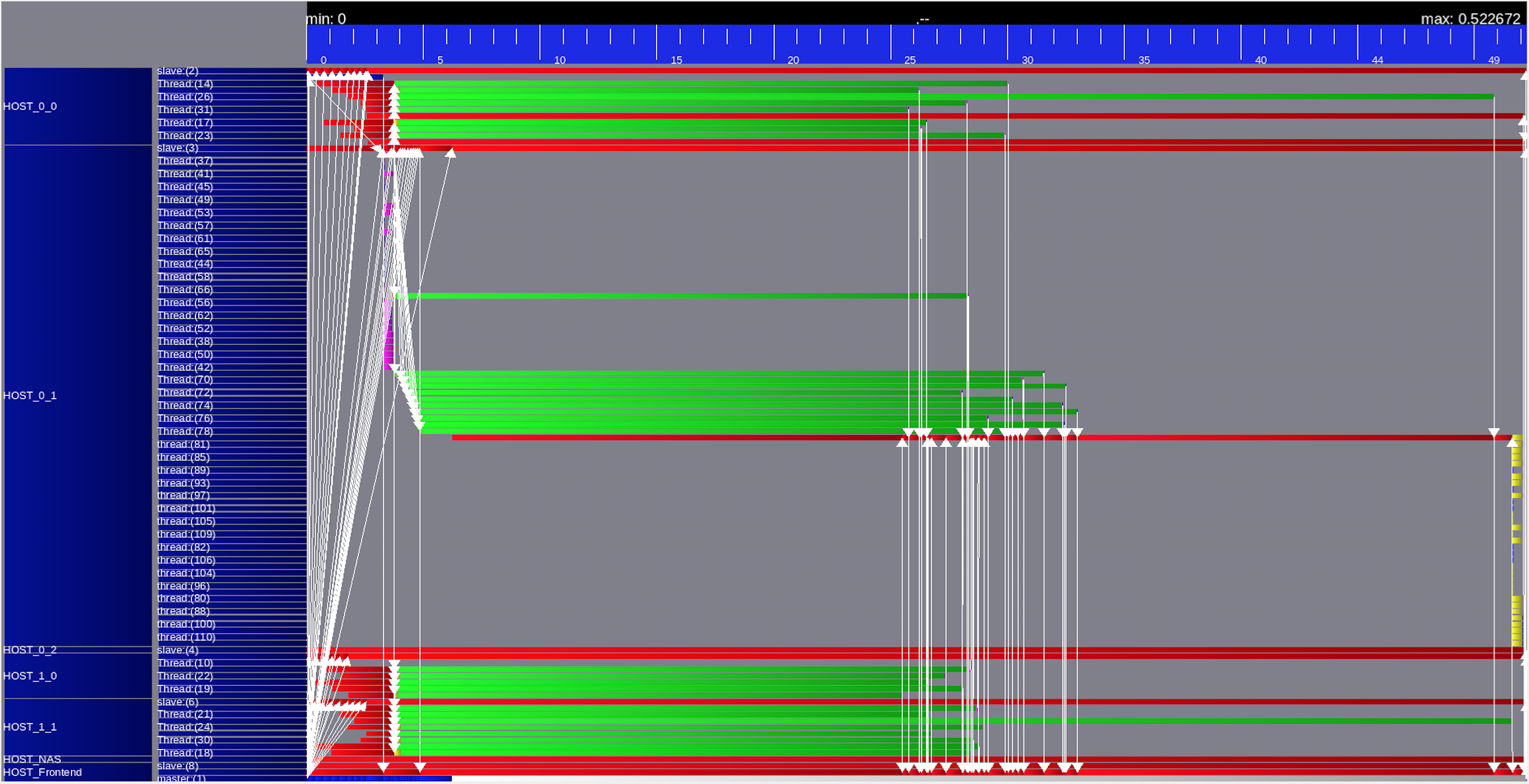}}} \qquad\qquad
    \subfigure[Worst mapping scenario]{\label{FIG:worst}\resizebox{150mm}{!}{\includegraphics{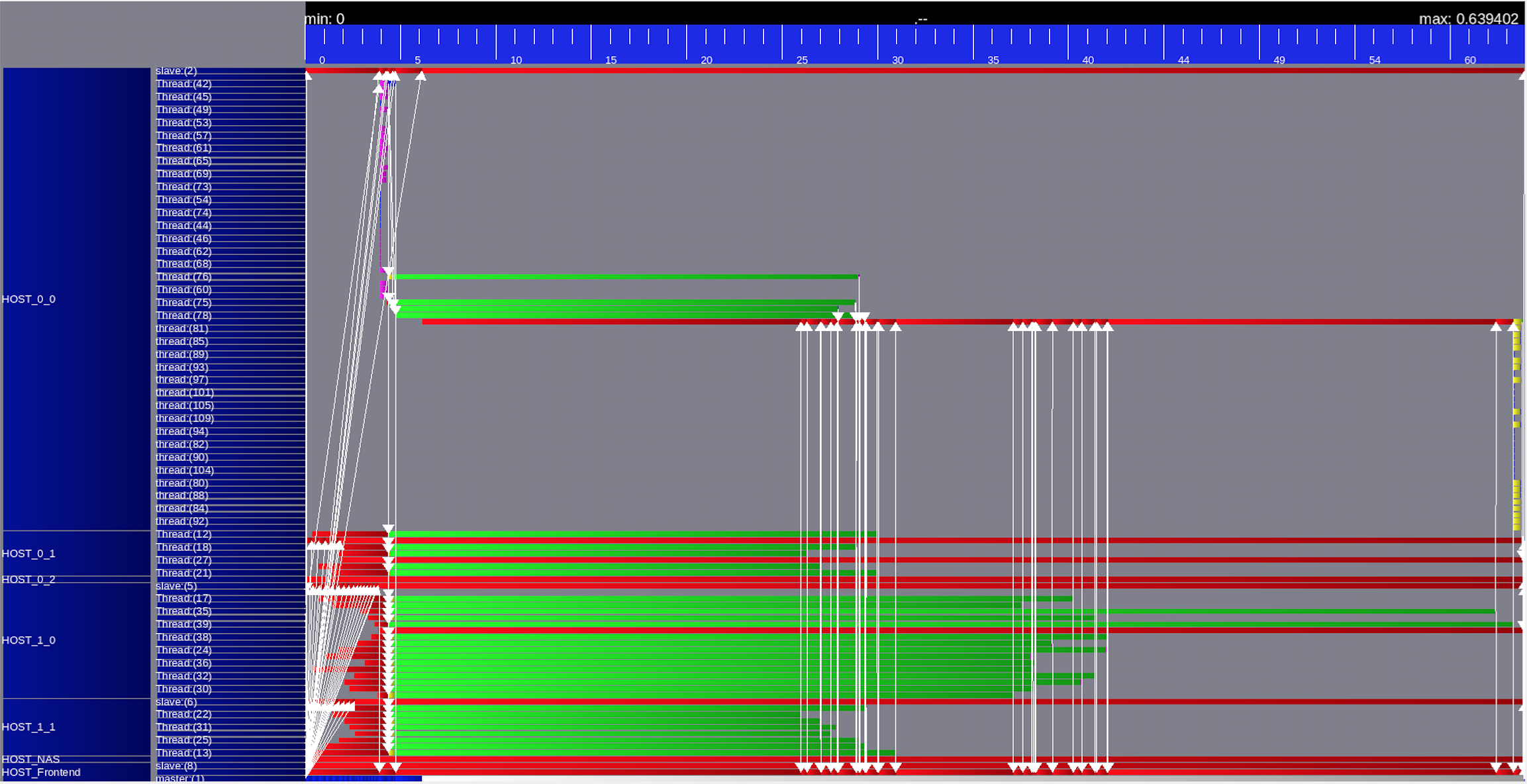}}}
  \end{center}
  \caption{Zoom into the best and worst scenarios in simulating the execution on the homogeneous platform}
  \label{FIG:scenarios}
\end{figure*}

The second experiment is similar to  the first one with the difference
that  the heterogeneous  platform  is  considered.  The  corresponding
results are  shown in  Figure~\ref{FIG:heterogeneous}.  Here,  for the
mappings with  the shortest and  longest execution time,  the detailed
temporal    evolution    of    the   execution    is    depicted    in
Figures~\ref{FIG:best}   and~\ref{FIG:worst}  respectively.    We  can
notice that in this heterogeneous  case, the mapping with the shortest
execution time may not be the most efficient one from the energy point
of view. Indeed, the mapping number 28  may be a good candidate, as it
provides  an execution  time close  to  the shortest  one while  using
around 15 percent less of energy.

\setlength{\textfloatsep}{5pt}

\begin{figure}[t]
  \begin{center}
    \subfigure[Execution time]{\label{FIG:LSgpu1}\resizebox{90mm}{!}{\includegraphics{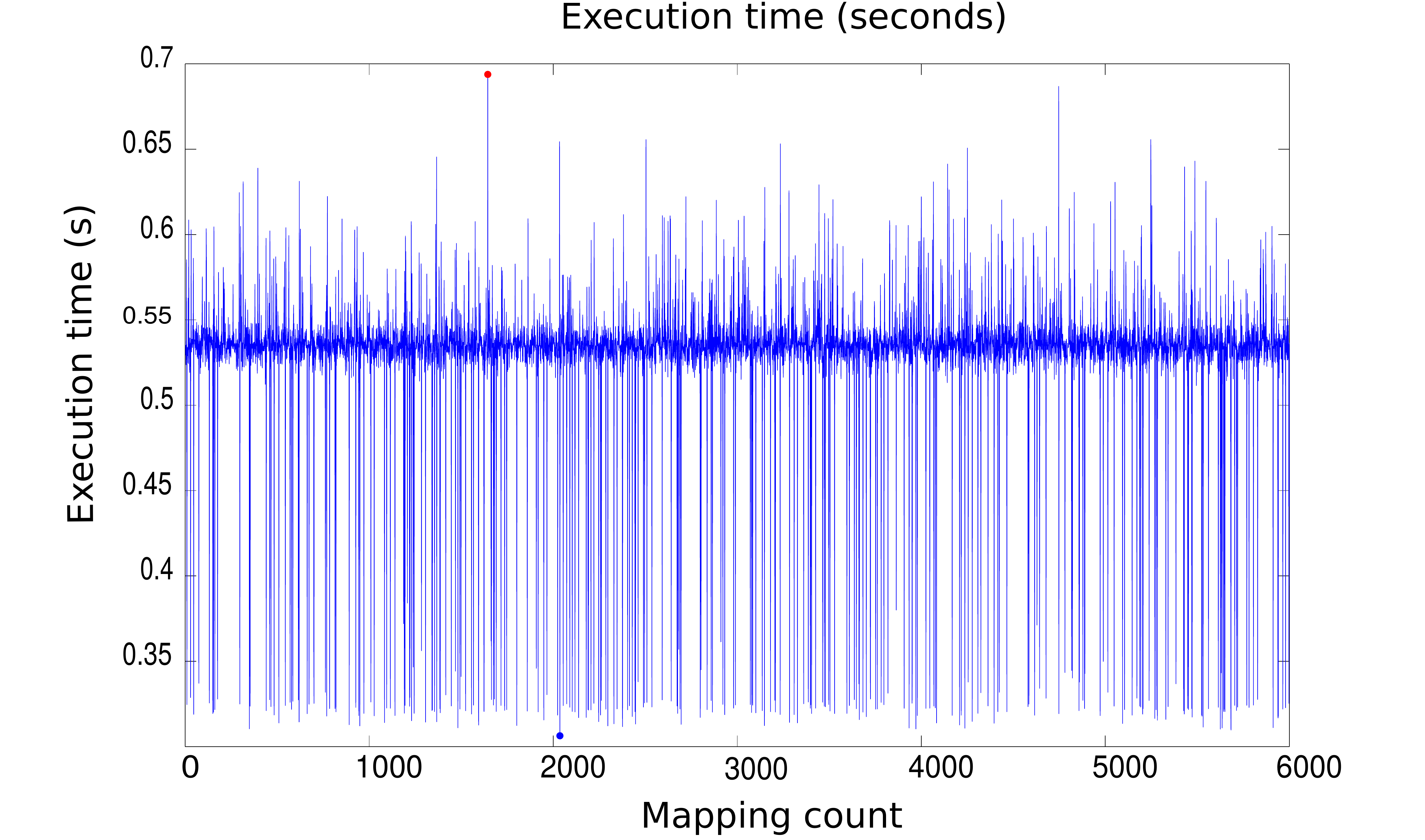}}} \qquad\qquad
    \subfigure[Energy consumption]{\label{FIG:LSgpu2}\resizebox{90mm}{!}{\includegraphics{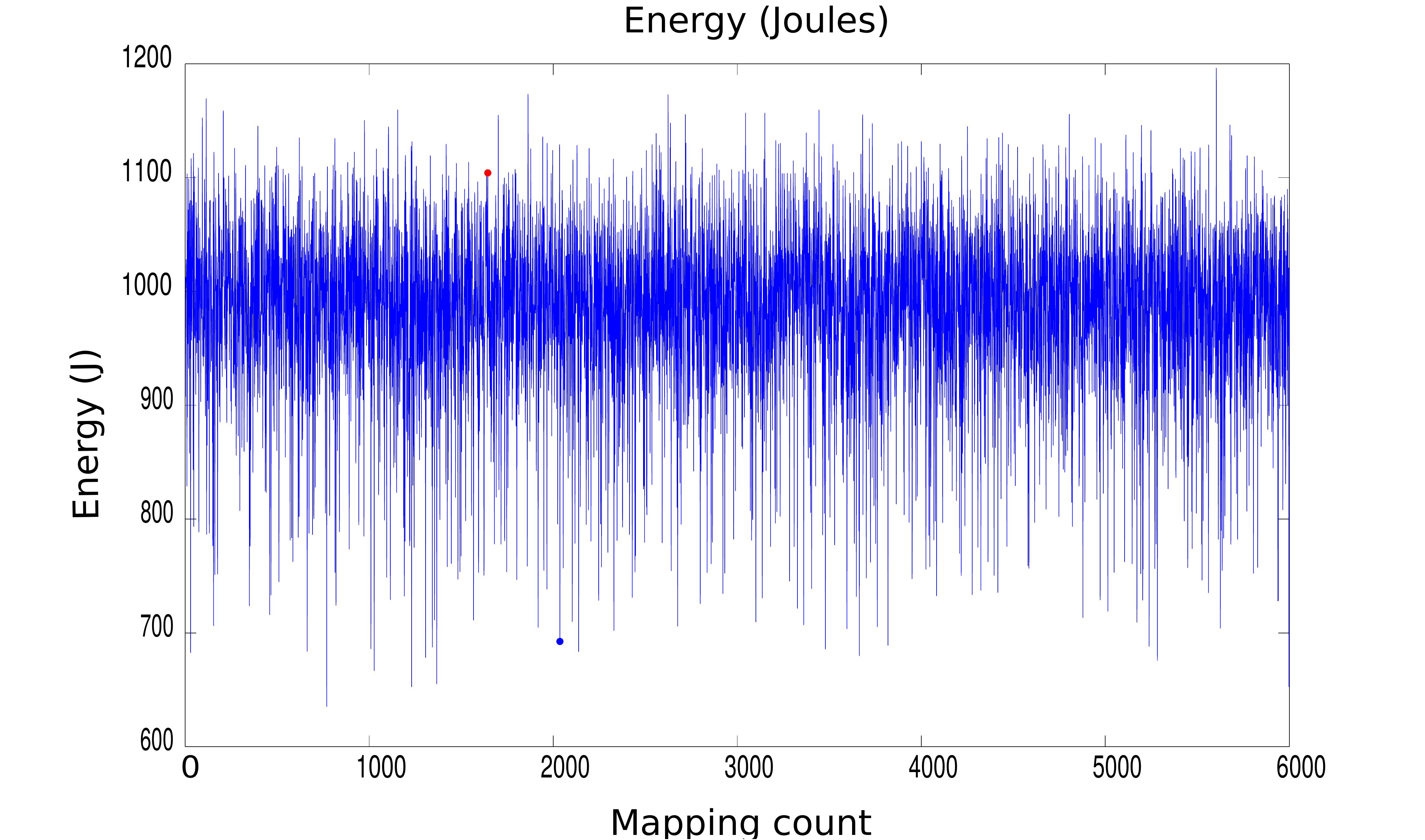}}} \qquad\qquad
    \subfigure[Simulation time]{\label{FIG:LSgpu3}\resizebox{90mm}{!}{\includegraphics{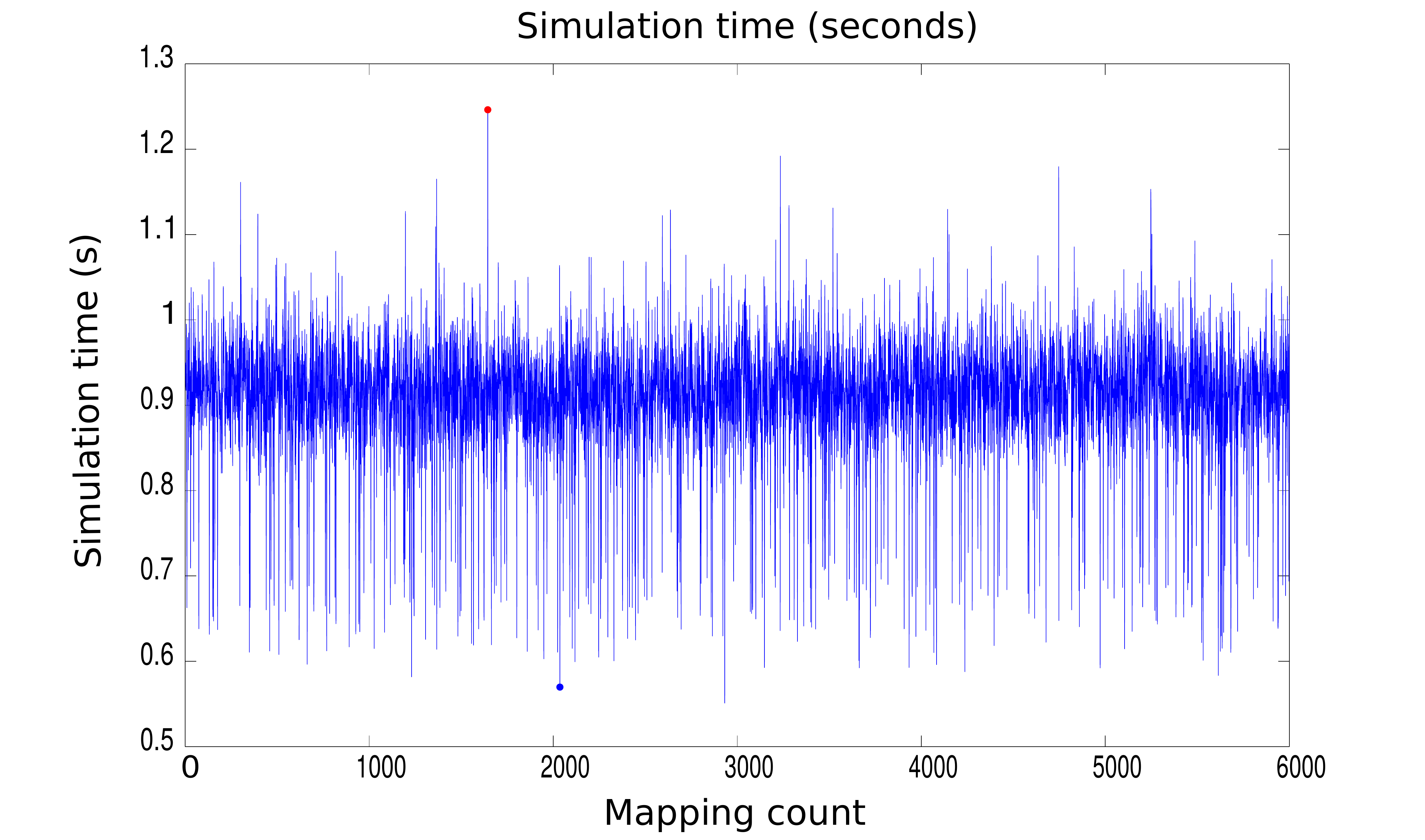}}}
  \end{center}
  \caption{Large scale simulation on heterogeneous platform}
  \label{FIG:Large_Scale}
\end{figure}

The  execution  time  in  Figure~\ref{FIG:best}  is  shorter  than  in
Figure~\ref{FIG:worst} because in the former case the tasks \emph{MS2}
are   more   evenly   distributed    on   the   hosts.    Indeed,   in
Figure~\ref{FIG:worst},  the host  HOST\_0\_1  executes 15  \emph{MS2}
tasks   and   the   host   HOST\_0\_0  only   4   tasks.    Thus,   in
Figure~\ref{FIG:worst} the  host HOST\_0\_1 is overloaded  compared to
the other  hosts.  In that  case, the  host HOST\_0\_1 has  to provide
computational  ressources   to  more  tasks  than   the  other  hosts.
Consequently,  the tasks  on the  host  HOST\_0\_1 need  more time  to
execute. Indeed, in Figure~\ref{FIG:worst}  the tasks executing on the
host HOST\_0\_1  need 50\% more time  to finish compared to  the tasks
executing on the other hosts.

\subsubsection{Large-scale mapping evaluation}

We extend  the previous evaluation to  a larger number of  mappings in
order  to get  a better  insight of  the mapping  impact on  execution
time.  Figure~\ref{FIG:Large_Scale}  shows   the  evaluation  of  6000
mappings of  the eScience  application on the  heterogeneous platform.
As for  the experiments  in the previous  section, these  mappings are
randomly choosen.

The shortest  execution time is  approximately 0.3064 seconds  and the
longest  execution  time  is  0.6938 seconds.   The  reason  for  this
difference is that  in the first case, 34.4\% of  the \emph{MS2} tasks
are  mapped on  the host  HOST\_0\_2 which  is almost  100 times  more
powerful than the other hosts. In  the second case, only 18.8\% of the
\emph{MS2}  tasks  are  mapped  on  the  host  HOST\_0\_2.   Thus  the
execution for  the first mapping  finishes sooner than for  the second
mapping. Table~\ref{table:1} shows the execution times (E\_time(s)) in
seconds of different  mappings. Each line corresponds  to one mapping.
For each  mapping, the  percentage of the  \emph{MS2} tasks  that were
mapped on the different hosts is specified. The mapping m\_good is the
best mapping  in terms of the  execution time among the  6000 mappings
depicted  in Figure~\ref{FIG:Large_Scale}.   The m\_bad  is the  worst
mapping among those  6000 mappings.  For the mapping  m\_best, all the
tasks of  the eScience  application were mapped  to the  most powerful
host HOST\_0\_2. In case of m\_worst, all the tasks were mapped on the
host HOST\_0\_1. When mapping all the tasks on the most powerful host,
the execution  time is almost 10  times shorter than if  all the tasks
are mapped on another host.   This result was expected.  Indeed, among
the 5 hosts of our platform, 4 hosts have identical performances and 1
host provides 100 times more flops compared to any other host.


\begin{table}[h!]
\centering
\begin{tabular}{c|c|c|c|c|c|c|l}
\cline{2-7}
& H\_0\_0 & H\_0\_1 & H\_0\_2 & H\_1\_0 & H\_1\_1 & E\_time(s)\\ \cline{1-7}
\multicolumn{1}{ |c| }{m\_bad} & 25\% & 12.5\% & 34.4\% & 9.4\% & 18.7\% & 0.6938   \\ \cline{1-7}
\multicolumn{1}{ |c| }{m\_good} & 53\% & 6.3\% & 18.8\% & 9.4\% & 12.5\% & 0.3064  \\ \cline{1-7}
\multicolumn{1}{ |c| }{m\_worst} & 0 & 100\% & 0 & 0 & 0 & 1.0270 \\ \cline{1-7}
\multicolumn{1}{ |c| }{m\_best} & 0 & 0 & 100\% & 0 & 0 & 0.1181 \\ \cline{1-7}
\end{tabular}
\caption{Execution times and \emph{MS2} tasks mappings for different mappings}
\label{table:1}
\end{table}

\section{Conclusion and Perspectives}
\label{SEC:Conclusions}

This  work  proposes an  integrated  simulation  workflow allowing  to
quickly  assess  the quality  of  mapping  algorithms targeting  cloud
infrastructures  and according  to different  criterion (i.e.   energy
efficiency  or execution  time).   This simulation  workflow takes  as
input  applications specifications  expressed  in  the \am{}  model
chosen  as the  application model  to be  used in  the context  of the
DreamCloud project. The specification of custom mappings is handled by
implementing a  clean interface  whose main role  is to  specify where
\am{}   entities  should   be   located  on   the  targeted   cloud
architecture.

We plan to  extend the mapping design space  exploration introduced in
the  last  section  with  optimization techniques  such  as  simulated
annealing to allow  end users to identify what are  (and why) the best
mapping strategies  for particular applications. We  will also improve
the  quality of  the  provided  results by  taking  into account  more
architectural parameters.

\section*{Acknowledgment}
The research  leading to these  results has received funding  from the
European Community's Seventh Framework Programme (FP7/2007-2013) under
the DreamCloud Project: \url{http://www.dreamcloud-project.org}, grant
agreement  no. 611411.   The  authors  would like  also  to thank  the
SimGrid team for the provided support.

\bibliographystyle{IEEEtran}
\bibliography{hipeac16}

\end{document}